# The Synthesis and Characterization of LiFeAs and NaFeAs

C.W.Chu[1,2,3], F. Chen[1], M. Gooch[1], A. M. Guloy[4], B. Lorenz[1], B. Lv[4], K. Sasmal[1], Z. J. Tang[4], J. H. Tapp[4] and Y. Y. Xue[1]

[1]Department of Physics and TCSUH, University of Houston, Houston, TX 77204-5002, USA

[2]Lawrence Berkeley National Laboratory, 1 Cyclotron Road, Berkeley, CA 94720, USA

[3]Hong Kong University of Science and Technology, Hong Kong, China

[4]Department of Chemistry and TCSUH, University of Houston, Houston, TX 77204-5002, USA

**Corresponding Author**

C. W. Chu

Texas Center for Superconductivity

University of Houston

202 Houston Science Center

Houston TX  77204-5002

PHONE:      713-743-8222

FAX:        713-743-8201

e-mail:     cwchu@uh.edu

**Abstract**

The newest homologous series of superconducting As-pnictides, LiFeAs (Li111) and NaFeAs (Na111) have been synthesized and investigated. Both crystallize with the layered tetragonal anti-PbFCl-type structure in P4/nmm space group. Polycrystalline samples and single-crystals of Li111 and Na111 display superconducting transitions at ~ 18 K and 12-25 K, respectively. No magnetic order has been found in either compound, although a weak magnetic background is clearly in evidence. The origin of the carriers and the stoichiometric compositions of Li111 and Na111 were explored.

**1. Introduction**

Throughout the history of superconductivity, the discovery of a new material system has often led to a paradigm shift in our understanding of the phenomenon and has provided new clues in the search for novel superconductors with a higher transition temperature ($T_c$). Previous examples include the A15 superconductors, the rare-earth chalcogenide ternaries, the rare-earth rhodium borides, the organic superconductors, the heavy Fermion superconductors and the cuprates, among others. The iron-pnictide system discovered early last year by Hosono et al. [1] is the most exciting example since the cuprates.

High temperature superconductors (HTSs) exhibit common features, such as a layered structure, atomic substructures, strong electron-electron interaction, instabilities and magnetic fluctuations, that may play important roles in the occurrence of high temperature superconductivity (HTS) [2]. There exists a large class of equiatomic quaternary layered compounds ROTPn, that assume the tetragonal ZrCuSiAs structure type, where R = rare-earth element, O = oxygen, T = transition-metal element, and Pn = pnictogen [3]. These consist of two layered substructures, namely, the alternating TPn-layers and the RO-layers. Each TPn-layer is composed of a T-sheet sandwiched by two Pn-sheets with the T-atoms tetrahedrally coordinated to four Pn-atoms. Each RO-layer consists of an O-sheet sandwiched between two R-sheets. TPn-layers may be expected to form the active block for the charge carriers to flow, and the RO-layers constitute the charge reservoir blocks that inject charge carriers into the active block to maintain the maximum layer-integrity of the TPn-layers, as in the cuprate HTSs. Differences do exist between the ROTPn and the cuprates, e.g. in the former, T is tetrahedrally coordinated with four Pn-atoms, whereas in the latter, Cu-atoms are coordinated by four O atoms to form a square plane.

Hosono et al. [4] first discovered superconductivity in LaOFeP with a $T_c \sim 4$ K. By increasing the carrier concentration through the partial replacement of O by F, its $T_c$ was raised to $\sim 9$ K. Shortly thereafter, LaONiP was found to exhibit a $T_c \sim 3$ K. However, immense excitement did not arise until the discovery of superconductivity at 26 K in the F-doped LaOFeAs (La1111), after the SDW was suppressed by electron-doping via the partial substitution of O by F in early 2008 [1]. The relatively high $T_c$ and the presence of a large concentration of the magnetic Fe, which is antithetic to superconductivity, in doped La1111 give us new hope in this novel material system to help achieve higher $T_c$ and to reveal the role of magnetism in HTS.

Electron-doped La1111 did not disappoint us. In the few weeks following the report of Honsono et al. in February 2008, the $T_c$ was raised quickly to 55 K by replacing La with rare-earth elements (R) of smaller radii in ROFeAs (R1111) [5] or by partial removal of O in R1111 [6]. Pressure was also found to raise the $T_c$ of La1111 at a high rate of +1.5 K/GPa [7]. Optimism concerning enhancing $T_c$ returned after the long $T_c$-stagnation since 1993. Extensive theoretical and experimental studies have since been carried out [8]. In the last six months, many layered Fe-pnitides have been found to be superconducting. By studying these compounds, the less critical role of O has been demonstrated, the importance of low dimensionality confirmed, the significant role of magnetic fluctuations in HTS critically reexamined, the electron-hole symmetry demonstrated, the origin of SDW ascertained, the less critical characteristic of layer-integrity (of the FeAs-layers) and thus the greater flexibility in doping in Fe-pnictides affirmed, and the absence of an insulator state shown. The layered Fe-pnictide superconductors discovered to date can be grouped into four homologous series, i.e. 1111 (ROFeAs [1] and AeFFeAs [9] with Ae = alkaline-earth element), 122 (AeFe$_2$As$_2$ [10] and AFe$_2$As$_2$ [11] with A = alkaline metal), 111 (AFeAs) [12-14] and 011 (FeSe) [15]. Among the above series, AFe$_2$As$_2$ [11] and AFeAs [12-14] become superconducting apparently without external doping and other undoped nonsuperconducting ones can be made superconducting by the application of pressures [16]. There exist clear differences between the maximum $T_c$'s of the different homologous series: ~ 55 K for 1111, ~ 38 K for 122, ~ 20 K for 111 and ~13 K for 011.

While many of the questions raised [17] concerning the Fe-pnictide superconductors soon after their discovery have been answered, some remain unanswered. For example, the following three have been asked: 1. How high can the $T_c$ of 1111 (or the known Fe-pnictides, more appropriately now) be raised? 2. How can superconductivity be induced in FeAs-layered compounds without external doping? 3. Is SDW a prerequisite for Fe-pnictides to become superconducting? Among the four homologous series, 111 is the newest, only discovered very recently. We therefore would like to present our results on Li111 and Na111 and discuss whether these two members can offer to help answer some questions about Fe-pnictide superconductors, although some preliminary results of Li111 have appeared elsewhere.

## 2. Experimental and Discussion

To address whether the $T_c$ of the 1111 series can be raised to beyond 26 K, soon after its discovery we examined the pressure influence on the $T_c$ of electron-doped Sm1111 (SmO$_{1-x}$F$_x$FeAs) at different x's [18]. We found that pressure enhances the $T_c$ of samples in the underdoped region where $dT_c/dx > 0$ and suppresses the $T_c$ of samples in the overdoped region where $dT_c/dx < 0$, similar to the holed-doped cuprate HTSs (Fig. 1). Following the universal $T_c$-x relationship proposed by Presland et al. [19] for the cuprates, we therefore concluded that the $T_c$ of the optimized R1111 compounds should be in the 50's K and independent of R. Therefore, we have decided to search for new FeAs-layered compounds of different structures to see whether $T_c$ can be raised to above 50's K. After finding superconducting K122 and Cs122 with $T_c$ = 3.7 K and 2.4 K [11], respectively, we succeeded in synthesizing and characterizing Li111 [12] and Na111 and found them superconducting without external doping with $T_c$ = 18 K and 12-25 K, respectively. Details will be discussed below:

## 2.1. LiFeAs (Li111)

Pure FeAs precursors were first synthesized from the reaction of high-purity Fe (pieces 99.999%) and As (lumps, 99.999%) in sealed quartz containers at 600-800 °C. Polycrystalline LiFeAs samples were then synthesized [12] through solid-state reaction of high-purity Li (ribbons, 99.99%) and FeAs precursors at high temperatures. Stoichiometric amounts of the starting materials were sealed in welded Nb tubes in an Ar atmosphere. All manipulations were performed within a purified Ar-filled glove box, with total $O_2$ and $H_2O$ levels < 0.1 ppm. The reaction charges were jacketed within evacuated and sealed quartz containers, heated to 750 °C at a rate of 5 °C/min and kept for 4 days at 750 °C. The reactant was then slowly cooled to 150 °C at a rate of 0.05 °C/min and then air-quenched to room temperature. Polycrystalline samples of Li111 so obtained are black with metallic luster and moderately sensitive to moist air. Single crystals of size of a few tenths of a millimeter were obtained by prolonged annealing at low temperature inside the sealed Nb tube filled with purified Ar. Elemental analyses on micro-single crystals and polycrystalline samples using an inductively coupled plasma/mass spectrometer (ICPM) gave a Li:Fe:As ratio of 1.00(3):1.03(2):1.00(1), representing a stoichiometric composition of LiFeAs. Phase purity and the lattice parameters of the polycrystalline samples were investigated by using a Panalytical X'pert diffractometer. The X-ray powder diffraction displayed no detectable impurity phases. The structure of Li111 was determined by XRD on a single crystal with dimensions of ~ 0.28 x 0.14 x 0.02 $mm^3$ mounted in a glass fiber under a stream of cold nitrogen gas at -58 °C, using a Siemens SMART diffractometer. Li111 crystallizes in a tetragonal structure (P4/nmm) of the anti- PbFCl-type with a = 3.7914(7) Å and c = 6.364(2) Å, as shown in Fig. 2. Li111 features $Fe_2As_2$ layers based on edge-sharing $FeAs_4$-tetrahedra or derived from As capping of the Fe-square nets, above and below each center of the Fe-squares. The Fe-As bond distance within the layer is 2.4204(4) Å; the nearest Fe-Fe distance is 2.6819(4) Å.

Various physical properties of polycrystalline and single crystal Li111 samples were determined with different techniques. They include the low frequency (19Hz) resistivity ($\rho$) of the polycrystalline samples using the standard four-lead technique with the LR700 resistance bridge; the dc magnetization (M) and the dc magnetic susceptibility ($\chi$) using the Quantum Design SQUID magnetometer; the ac magnetic susceptibility ($\chi_{ac}$) at 19 Hz by the inductance technique using the Linear Research LR700 Inductance Bridge; the specific heat ($C_P$) employing the relaxation technique with the Quantum Design PPMS; and the thermoelectric power (S) employing a high precision low frequency technique developed by us. The hydrostatic pressure environment up 1.8 GPa was measured inductively using the clamp technique with the non-magnetic Be-Cu high pressure cell, where the pressure was measured by a superconducting Pb-manometer placed next to the sample.

Typical temperature dependences of $\rho(T)$ and $\chi(T)$ of polycrystalline Li111 samples are shown in Fig. 3. The $\rho(T)$ decreases with temperature decrease as a metal but with a large negative curvature, indicating the presence of strong electron-electron interactions in the compound. It drops rapidly to close to zero, indicative of a superconducting transition, with a $T_c \sim$ 18 K. This is further evidenced by the appearance of a diamagnetic shift in $\chi(T)$ at ~ 18 K, corresponding to ~ 100% of a bulk superconductor. The small but non-zero residual $\rho$ is attributed to the nonconducting grain surfaces resulting from the chemical sensitivity of the polycrystalline sample to moist air. S(T), which is known to be insensitive to the interface resistance of grain boundaries, indeed confirms a superconducting transition to zero at ~ 18 K. It also gives a negative value at room temperature, indicating an average electron behavior of the carriers. Upon application of a magnetic field, the transition is shifted downward as expected. By taking the temperatures where ρ drops by 5% of the transition as $T_c$ at different field, an upper critical field at zero temperature $H_{c2}(0) > 80$ T is obtained according to the Ginzburg-Landau formula of $H_{c2}(t) = H_{c2}[(1-t^2)/(1+t^2)]$, where $t = T/T_c$. The $C_p(T)$ in ambient and 7 T fields is shown in Fig. 4. A clear anomaly appears at ~ 16 K, indicating an unambiguous bulk superconductor. Under 7 T, the anomaly is broadened, reduced and pushed to lower temperatures. Analysis of the data between 10-20 K gives a Sommerfeld coefficient $\gamma = 22.5$ mJ/mol.K$^2$.

The pressure effect on $\chi_{ac}(T)$ of Li111 is shown in Fig. 5. The superconducting transition is clearly shifted downward in parallel by pressure at a rate of -1.56 K/GPa and the pressure-dependence of the onset-$T_c$ is summarized in the same figure.

Prior to the discovery of superconductivity in Li111, LDF calculations with virtual crystal approximation were carried out and it was shown by Singh [20] that the compound possesses a small Fermi surface with hole-cylinders at the zone-center and electron-cylinders at the zone corners, as well as high density of states N(E) below the Fermi surface $E_F$, similar to La1111. He predicted that Li111 should have an antiferromagnetic ground state and that N($E_F$), which comes mainly from the Fe 3d electrons that are insensitive to doping or Li-content. Indeed, our XRD discussed above shows that Li111 has a structure closely related to the undoped La1111. Without doping, Li111 has Fe$^{+2}$-ions like La1111, which is not superconducting and undergoes a SDW transition at 135 K as evidenced by a large ρ-drop [1]. In contrast to the above predictions, we found Li111 to be superconducting without evidence of a magnetic transition. This was also observed by other independent groups at about the same time [12-14]. Mössbauer measurements of Li111 have not found any indication of magnetic order at low temperatures [21]. According to later LDA calculations by Nebrasov et al. [22], one may attribute the absence of SDW transition or the weaker magnetic effect in Li111 to its smaller $N(E_F)$ resulting from the tighter As-tetrahedral coordination. Jishi and Alyahhyei [23] recently calculated the electron-phonon coupling constant of Li111 and found it to be too small to account for the relatively high $T_c$ of the compound. Both the XRD and the powder neutron diffraction (PND) data showed that Li111 should have the stoichiometric composition ratio of 1:1:1. Its presence of superconductivity and absence of SDW without external pressure or doping remain puzzling based on the valence count of Li111. Previous investigation by Juza and Langer [24] showed that LiFeAs could form only in its Li-rich and Fe-rich forms. If this is true, the dilemma concerning superconductivity and magnetism can then be resolved. Such a scenario may not be impossible, given the uncertainty of XRD and PND in determining light elements like Li. The negative S detected at room temperature indicates that the carriers in Li111 behave as electrons on the average, suggesting that the sample we have investigated may be Li-rich.

It has been demonstrated that for hole-doped cuprates and Fe-pnictides, $dT_c/dP$ is positive in the underdoped region where $dT_c/dx$ is positive, while $dT_c/dP$ is negative in the overdoped region where $dT_c/dx$ is negative (Fig. 1). This suggests that pressure produces the same effect as hole-addition or electron-reduction [17]. The observed negative $dT_c/dP$ in Li111 implies that the compound may be in the hole over-doped region or the electron underdoped region, consistent with the Li-rich state of the sample suggested.

## 2.2. NaFeAs (Na111)

Polycrystalline NaFeAs samples with nominal composition $Na_xFeAs$ with $0.5 \leq x \leq 1$ were synthesized by the solid-state reaction technique employed for Li111 [12], but with a small (5%) excess of Na. The Na111 polycrystalline samples are black with metallic luster. Post annealing around 300-400 °C, both inside the Ar-filled Nb tubes and under continuous vacuum, were also carried out for some synthesized samples. Plate-like single crystals with dimensions of a few tenths of a millimeter were also obtained after prolonged heat treatment at low temperature. All Na111 samples and the polycrystalline ones so obtained are extremely sensitive to air. Careful steps in handling the samples were taken to minimize the sample exposure to air. Both powder and single-crystal XRD analyses were performed, and the resistivity ($\rho$), the dc magnetization (M), the dc magnetic susceptibility ($\chi$), the thermoelectric power and the specific heat ($C_p$) were measured, by employing the same techniques used for Li111, described above.

The powder XRD data show that samples with nominal compositions $Na_xFeAs$, x = 1.0 and 0.9 are single-phase and all diffraction peaks can be indexed to a tetragonal unit cell with the P4/nmm space group. X-ray diffraction refinements were performed on data obtained from a single crystal with dimensions of ~ 0.08 x 0.06 x 0.04 mm mounted in a glass fiber under a stream of cold nitrogen gas at -57 °C. The XRD data were collected using a Siemens SMART diffractometer, using the Mo Kα1 radiation. Final cell parameters are: $a$ = 3.9866(16) Å, $c$ = 7.094(4) Å, V = 112.74(9) Å$^3$ Z = 2. Structure refinements indicate that Na111 is isostructural with Li111, crystallizing in the anti-PbFCl-type with a = 3.7914(7) Å and c = 6.364(2) Å, as shown in Fig. 6. This is in agreement with a previous report [13]. The Na111 structure also features $Fe_2As_2$ layers based on edge-sharing $FeAs_4$ tetrahedra. The Fe-As bond distance within the layer is 2.457(1) Å; the nearest Fe-Fe distance is 2.819(1) Å. The critical As-Fe-As bond angles are 108.44(4)° and 109.99(2)°, comparable to the As-Fe-As angles of 103.11(2)° and 112.74(1)° in Li111.

The $\rho(T)$ is shown in Fig. 7 for Na111 with nominal x = 0.9 (for reasons to be evident later). On cooling from 300 K, $\rho(T)$ decreases almost linearly with temperature decrease, deviates from this linearity below ~ 100 K, decreases more rapidly below ~ 50 K and finally drops to zero at 12 K, slightly higher than the $T_c$ previously reported [13]. The transition appears to be rather broad. When a field up to 7 T is applied, the overall transition is shifted progressively almost in parallel to lower temperature and the temperature where $\rho(T)$ becomes zero decreases as expected of a superconducting transition (Fig. 7). The onset $T_c$ is rather difficult to define from Fig. 7, although the data gives a value much higher than 12 K.

The $\chi(T)$ of a single crystal of Na111 shown in Fig. 8 exhibits a broad superconducting transition with a $T_c$ ~ 18 K and a large hysteresis, implying a large flux pinning force and possibly some magnetic background. Whether this magnetic background is intrinsic to the sample is not clear at this time. It is interesting to note, however, that the field-cooled and zero-field-cooled $\chi(T)$'s converge on warming at ~ 25 K, representing a higher onset $T_c$ near this temperature. The signal size of the zero-field-cooled diamagnetic shift at 4 K gives a superconducting volume < 10% of a bulk superconductor. Unfortunately, other single crystals of Na111 exhibit only a much smaller superconducting signal, if at all. Thus, we decided to examine the Na content (x)-dependence of the $\chi(T)$ of $Na_xFeAs$.

The magnetization of the polycrystalline samples with $0.5 \leq x \leq 1.0$ sealed either in thin quartz tubes or vacuum-grease coated gel-caps was measured at 10 Oe. Small but noticeable magnetic background appears, which has the lowest value of $4\pi\chi \approx 2\cdot10^{-3}$ at $x \geq 0.9$ and raises up to 0.1 for smaller x or after aging, including exposure to air. This background, fortunately, is only weakly dependent on the temperature between 20 and 50 K. Therefore, $4\pi(\chi-\chi_{30\,K})$ is used to extract information about the transition. The zero-field-cooled susceptibility $4\pi(\chi_{ZFC}-\chi_{ZFC,30\,K})$ and the field-cooled one $4\pi(\chi_{FC}-\chi_{FC,30\,K})$ appear to converge at temperature up to 40 K for the x = 0.5 sample, suggesting a possible onset $T_c$ at this temperature, consistent with the suggestion from the $\rho(T)$ results in Fig. 7. As shown in Fig. 9, the diamagnetic transition is rather broad, and the exact onset $T_c$ remains difficult to determine. The magnitude of the diamagnetic shift depends on x and demonstrates a 100% superconducting volume fraction only for the x = 0.9 sample (Fig. 9b). The rather small $4\pi(\chi_{ZFC}-\chi_{ZFC,30\,K})$ at the nominal x = 1.0, on the other hand, indicates that bulk superconductivity may not occur there (Fig. 9a).

In order to determine the exact x for the superconducting Na111 phase, we analyze the XRD and magnetization data and determine the x-dependences of $T_c$ and the superconducting volume fraction of Na111 of different x's. Detailed analyses will be published elsewhere. Here we shall report the preliminary results. Samples with x = 1.0 and 0.9 are pure phase of Na111 with the tetragonal structure of the anti-PbFCl-type, within the resolution of XRD (Fig. 6). The a-axis shrinks and the c-axis remains constant with a decrease in x. For samples with x < 0.9, noticeable impurity phases appear while the majority phase retains the Na111 structure and the lattice parameters of the majority phase vary negligibly with x. By examining the amount of phases present in the samples and by balancing the masses of the constituents, the actual Na-value of the Na111 phase in the samples with x < 0.9 is close to 0.86 for the x = 0.8 and 0.7 samples. For the x = 0.5 sample, the presence of a large amount of impurity phases make the analysis less reliable.

As described above, magnetic measurements carried out at 10 Oe show small but noticeable magnetic background which has the lowest value of $4\pi\chi \approx 2\cdot10^{-3}$ at y ≥ 0.9 and rises up to 0.1 as x decreases or after air exposure. This background, fortunately, depends only weakly on temperature between 20 and 50 K. The almost same $4\pi(\chi_{FC}-\chi_{FC,30\,K})$ and $4\pi(\chi_{ZFC}-\chi_{ZFC,30\,K})$ at x ~ 0.9 indicate that samples of this x have relatively weak pinning. The isothermal M-H loops around 4 K << $T_c$ support the interpretation that the flux pinning is rather weak over the x-range so the deduced $4\pi(\chi - \chi_{30\,K})$ in the field-cooling can be taken as a measure of the relative superconducting volume fraction in the samples. They are shown in Fig. 9. The sharpness of the x-dependence of $4\pi(\chi_{FC}-\chi_{FC,30\,K})$ on x and its large value at x = 0.9 suggests that bulk superconductivity takes place in Na-deficient Na111-phase with a narrow x-range.

To understand the observed superconductivity in the stoichiometric Na111, we further examined the "aging" effect on these samples. An aging effect on the samples with x = 1.0 is clearly evident in Fig. 10. Both the $T_c$ and the superconducting signal grow when the sample is aged, which occurs even in a vacuum-grease sealed gel-cup. The effect can be understood in terms of the loss of Na when it reacts with whatever small amount of $O_2$ or $H_2O$ in the sample environment. All these provide another piece of evidence that superconductivity occurs in Na-deficient Na111. This is also consistent with the electron-doping characteristic of the carriers of the superconducting samples as indicated by the negative S(T) shown in Fig. 11.

## 3. Conclusion

We have synthesized and characterized both polycrystalline and small single-crystal Li111 and Na111. Both crystallize in a layered tetragonal anti-PbFCl-type structure with a space group of P4/nmm. The lattice parameters (a,c) are (3.7914 Å, 6.364 Å) for Li111 and (3.95 Å, 7.04 Å) for Na111, respectively. In contrast to conventional wisdom, they are found to be superconducting with a $T_c$ ~ 18 K for Li111 and 12-18 K for Na111 without external doping, although superconductivity occurs in Na-deficient samples over a very narrow x-range close to 0.9. The onset $T_c$ of Na111 may be even as high as 25 K. Carriers in both compounds exhibit an electron-behavior.

The compounds of Li111 and Na111 add a new homologous series 111 series to the superconducting Fe-pnictide system. Together with the previously known 1111, 122, and 011 series, 111 may provide additional insights to the role of magnetism in the occurrence of HTS, given its lack of it, at least in terms of magnetic ordering. The four homologous series do offer basic ideas concerning the formation of compounds with more complex structures for higher $T_c$. Unfortunately, the similarities between them overwhelm the difference. Given the large number of layer transition-metal-pnictides, the system as a whole still holds promise for higher $T_c$. As to the two new members of the 111 series, two important issues warrant further careful investigation, i.e. the exact compositional ranges and the origin of the magnetic background in these compounds, although preliminary investigations have been done.

**Acknowledgments**

This work is supported in part by the T.L.L. Temple Foundation, the John J. and Rebecca Moores Endowment, the State of Texas through TCSUH, the U.S. Air Force Office of Scientific Research, and the LBNL through the U.S. Department of Energy. A.M.G. and B.L. acknowledge the support from the NSF (CHE-0616805) and the Robert A. Welch Foundation.

**Figure Captions:**

Figure 1: $T_c$ *vs* doping phase diagram of high temperature superconductors (schematic).

Figure 2: Crystal structure of LiFeAs.

Figure 3: Resistivity $\rho(T)$ and volume susceptibility $\chi(T)$ (inset) of LiFeAs.

Figure 4: Heat capacity $C_p(T)$ of LiFeAs at zero magnetic field and at H=7 Tesla.

Figure 5: $T_c$ *vs* pressure of LiFeAs. Inset: $\chi_{ac}$ *vs* T at different pressures.

Figure 6: X-ray diffraction spectrum of a NaFeAs (x = 1.0) polycrystalline sample.

Figure 7: Resistance R(T) of a $Na_{0.9}FeAs$ sample in zero field. Inset: $\rho(T)$ at different fields. From left to right: 7, 5, 3, 1 and 0 T.

Figure 8: Volume susceptibility χ(T) of a NaFeAs (x = 1.0) single crystal. Triangles: $M_{FC}$ and circles: $M_{ZFC}$.

Figure 9: Diamagnetic shifts *vs* T for $Na_xFeAs$ with (a) x = 1.0, (b) x = 0.9, and (c) x = 0.5, respectively. The open and filled symbols are the field-cooled and zero-field-cooled susceptibilities, respectively.

Figure 10: χ *vs* T for an aged NaFeAs (x = 1.0) sample. The top and bottom figures are for the zero-field-cooled and field-cooled values of χ, respectively. Circles: immediately measured; triangles: 8 hrs later; and squares: 66 hrs later.

Figure 11: Thermoelectric power S(T) of a NaFeAs (x = 1.0) sample.

Figure 1:

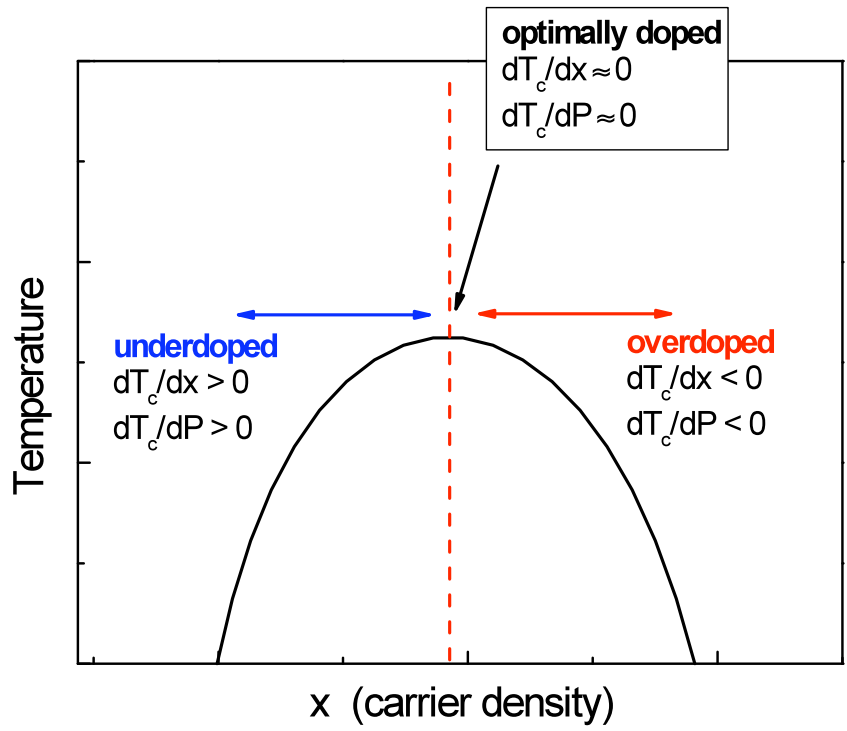

Figure 2:

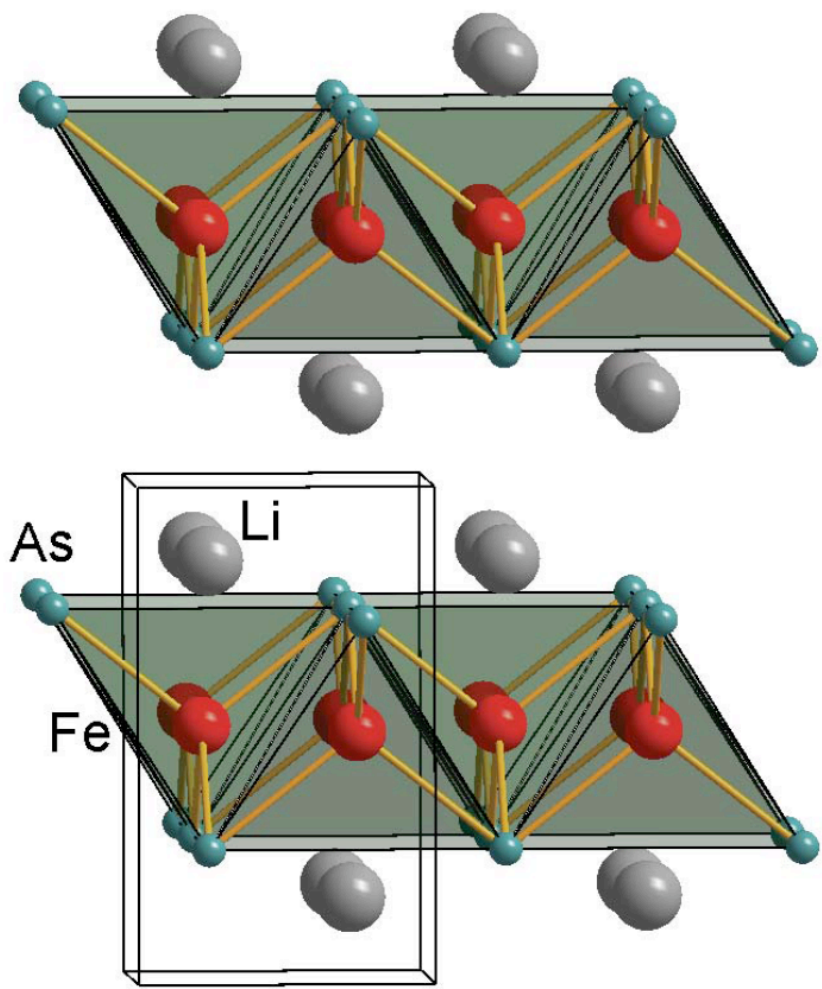

Figure 3:

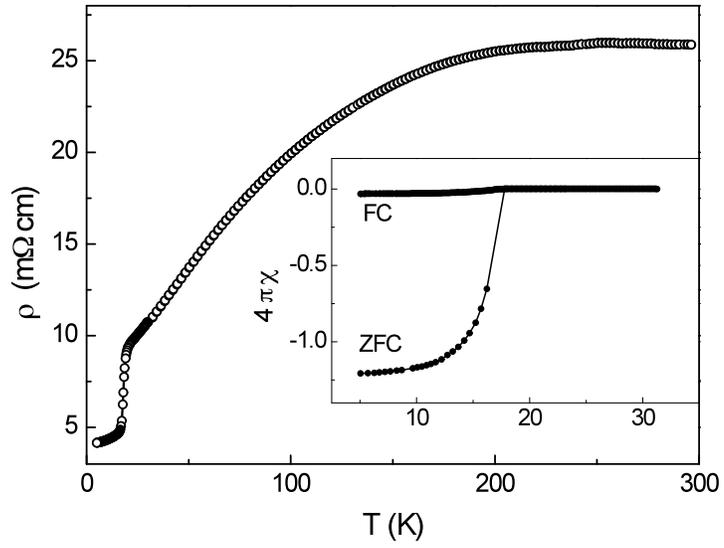

Figure 4:

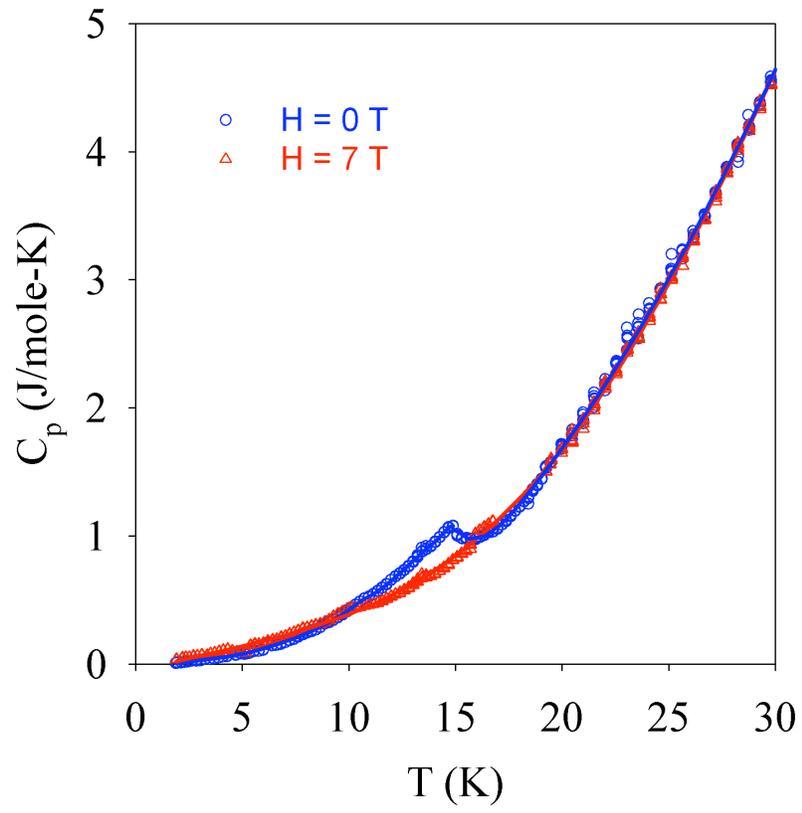

Figure 5:

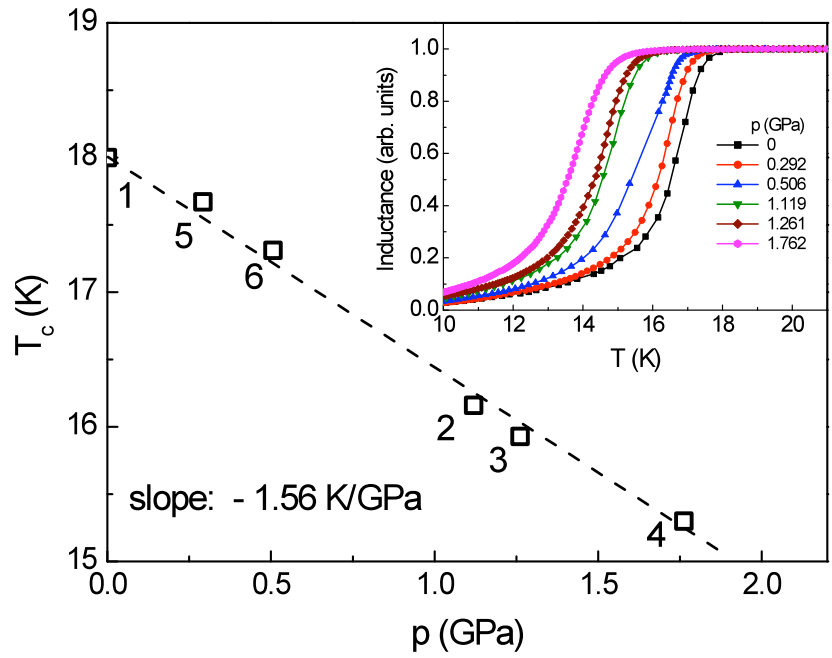

Figure 6:

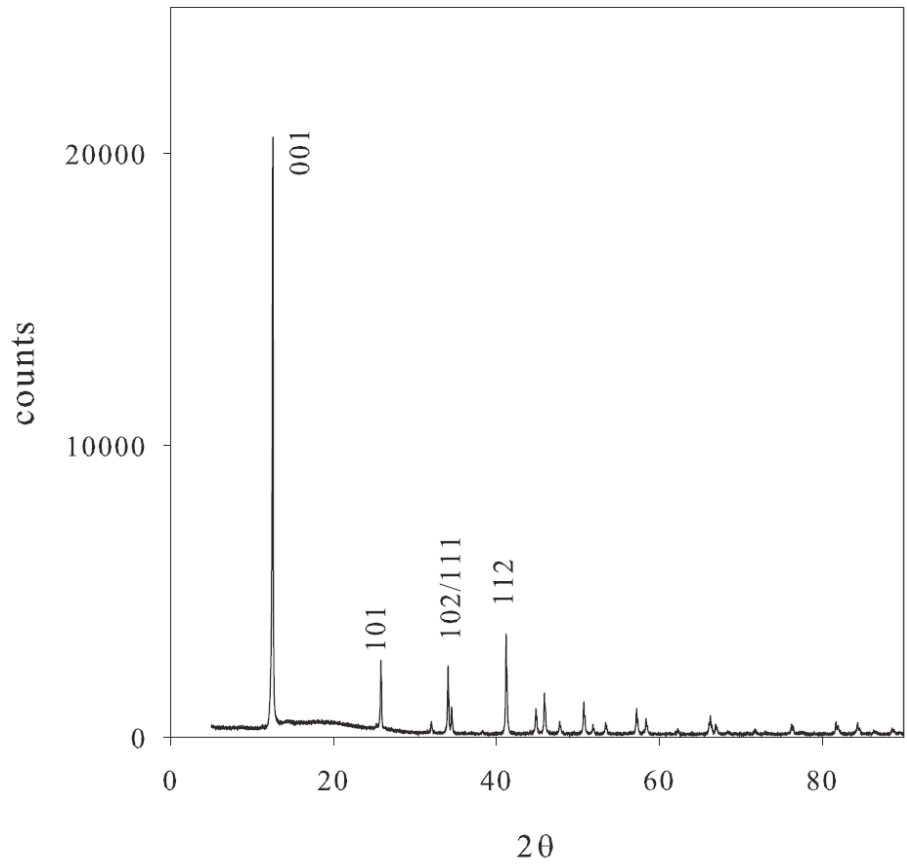

Figure 7:

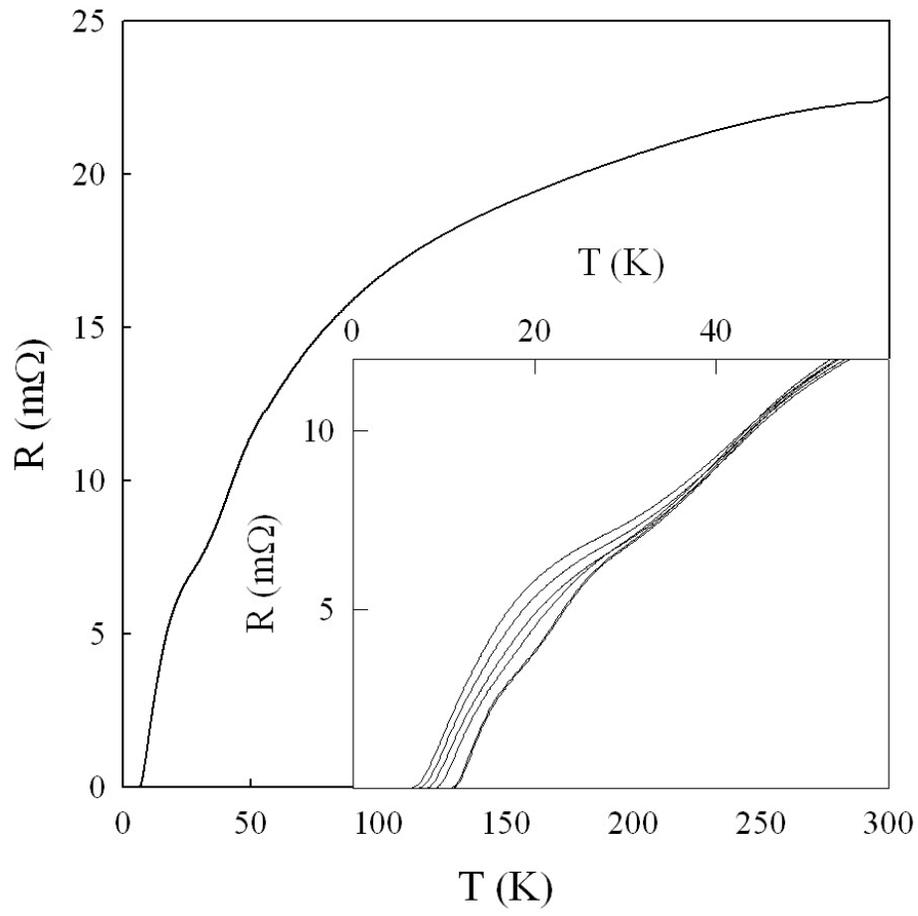

Figure 8:

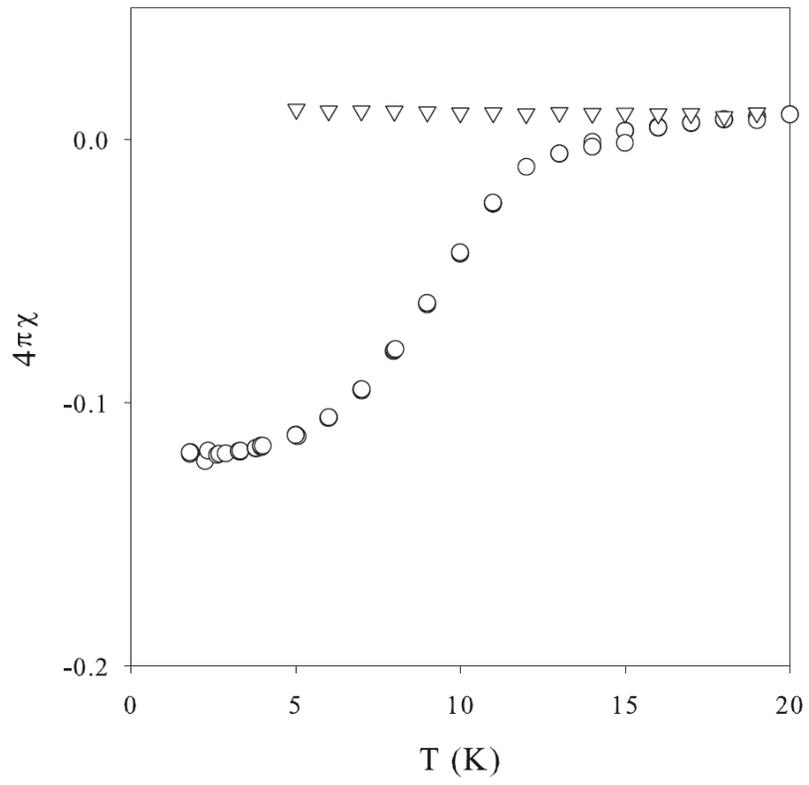

Figure 9:

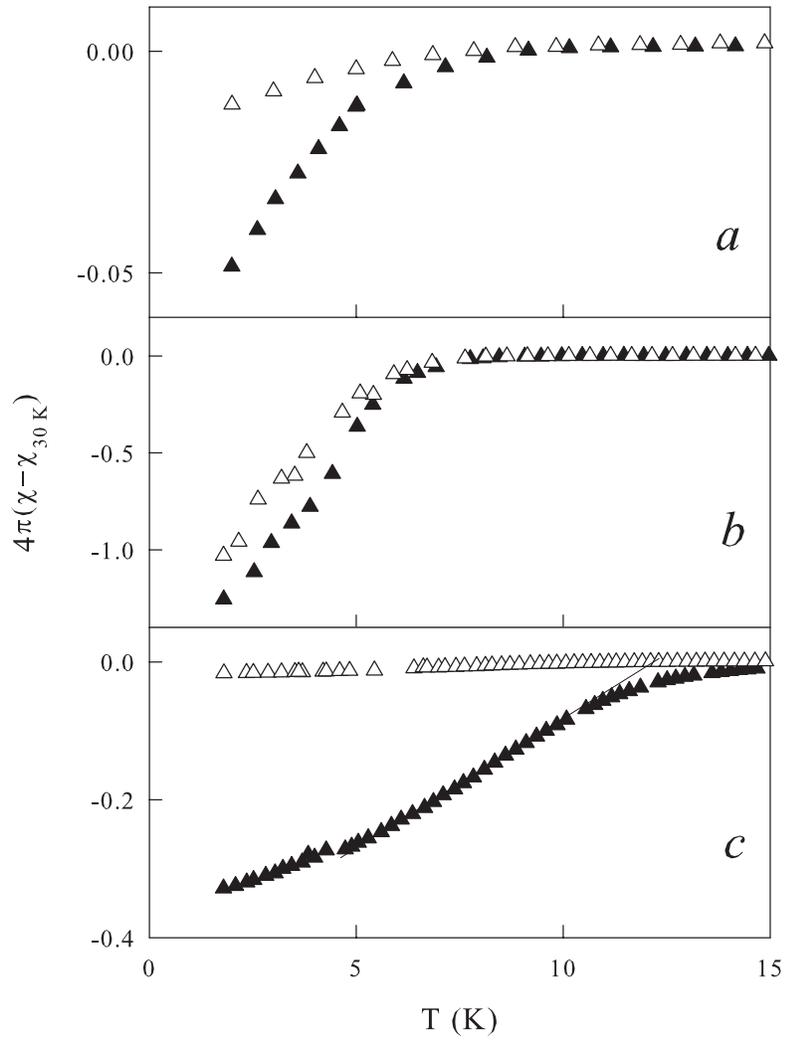

Figure 10:

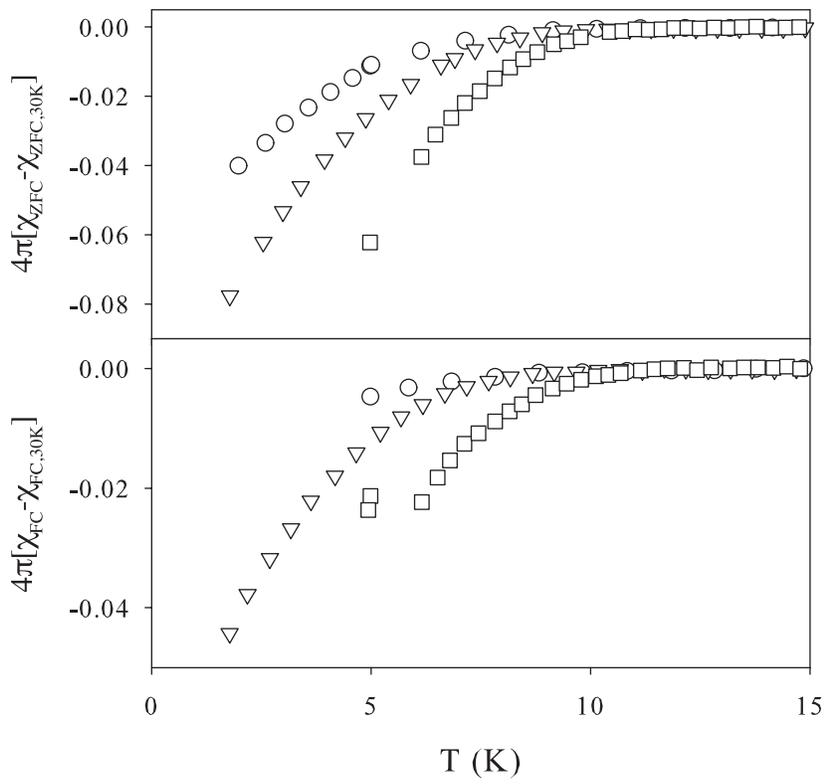

Figure 11:

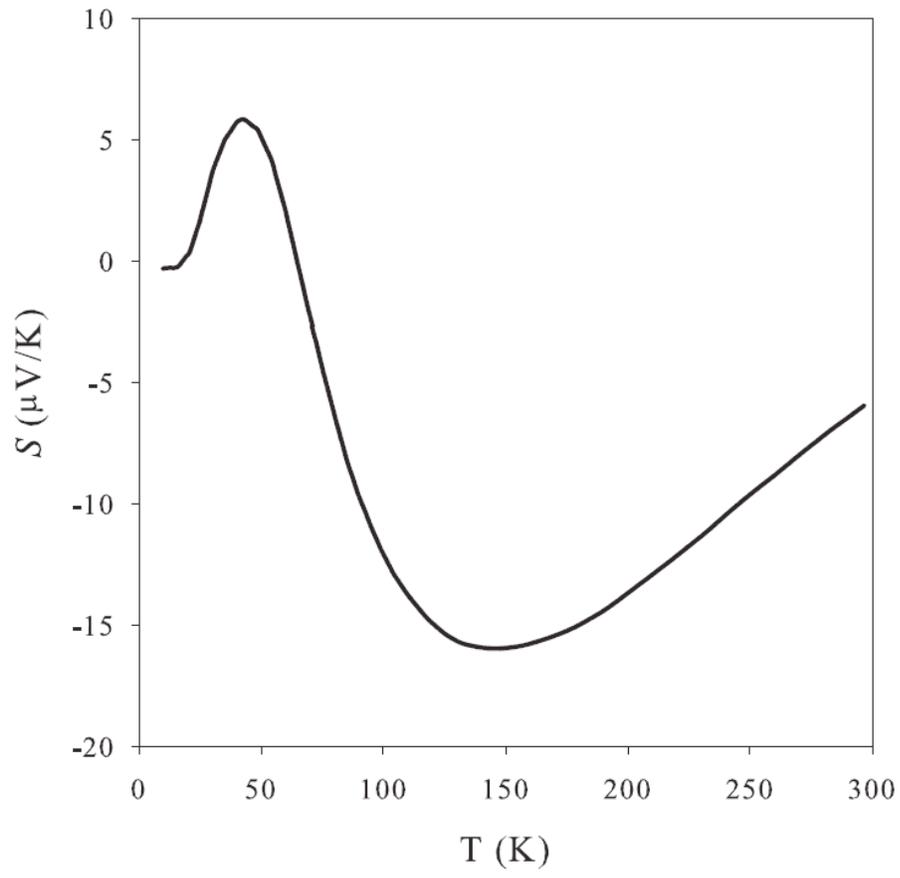